\documentclass{revtex4}
\usepackage{amssymb}
\usepackage{amsfonts}
\usepackage{amsmath}

\input{tcilatex}

\begin{document}

\preprint{}
\title{Regular charged black hole construction in $2+1$ $-$dimensions}
\author{S. Habib Mazharimousavi}
\email{habib.mazhari@emu.edu.tr}
\author{M. Halilsoy }
\email{mustafa.halilsoy@emu.edu.tr }
\author{T. Tahamtan}
\email{tayabeh.tahamtan@emu.edu.tr}
\affiliation{Department of Physics, Eastern Mediterranean University, G. Magusa, north
Cyprus, Mersin 10 - Turkey, Tel.: +90 392 6301067; fax: +90 3692 365 1604.}
\keywords{Regular charged black; BTZ black hole; Nnonlinear electrodynamics;
Exact solution}

\begin{abstract}
\textbf{Abstract:} It is well-known that unlike its chargeless version the
charged Banados-Teitelboim-Zanelli (BTZ) black hole solution in $2+1-$
dimensional spacetime is singular. We construct a charged, regular extension
of the BTZ black hole solution by employing nonlinear Born-Infeld
electrodynamics, supplemented with the Hoffmann term and gluing different
spacetimes. The role of the latter term is to divide spacetime in a natural
way into two regions by a circle and eliminate the inner singularity.
Thermodynamics of such a black hole is investigated by Kaluza-Klein
reduction to the 1+1-dimensional dilaton gravity.
\end{abstract}

\pacs{PACS number}
\maketitle

\section{Introduction}

The principal aim in introducing nonlinear electrodynamics was to eliminate
the Coulomb divergences that arose in Maxwell electromagnetism. One such
prominent member of this class of theories was introduced by Born and Infeld
(BI) in 1934 \cite{1}. The characteristic action of the BI theory consisted
of a square root term under which the electromagnetic field strengths
admitted automatic upper (lower) bounds. This is reminiscent of the
relativistic particle Lagrangian where the upper bound turns out to be the
speed of light. In the proper limiting procedure one naturally recovers from
the BI formalism the linear electromagnetism of Maxwell as expected.

One problem that was invited with the BI theory was the double-valued
behavior of the displacement vector $\overrightarrow{D}(\overrightarrow{E})$
as a function of the electric field $\overrightarrow{E}$ \cite{2}. That
means, with the same $\overrightarrow{E}$ value at a particular point one
could obtain two different values for $\overrightarrow{D}$\ ; a totally
unacceptable situation from the physics standpoint. To resolve this problem,
shortly after the introduction of BI theory, Hoffmann and Infeld introduced
a supplementary term to the BI Lagrangian \cite{2,3} which came to be known
as the 'Hoffmann term'. This amounted to a pair of Lagrangians matching at a
natural boundary, removing the double-values in $\overrightarrow{D}(%
\overrightarrow{E})$\ , but all at the cost of two Lagrangians in the same
theory.

These novel ideas of 1930s may find appropriate arena now in the geometric
theory of Einstein, namely the general relativity since boundaries /
intersections of different spacetimes are encountered therein naturally.
From this token we developed a geometric model of a particle where the
outside and inside of the particle correspond to different spacetimes
matching consistently at the radius of the particle \cite{4,5}. This took
place in the $4-$ dimensional $\left( d=4\right) $ spacetime, but naturally
we can extend the investigation also to nowadays fashionable $d>4$
spacetimes \cite{6}. In this paper instead of $d>4$\ we shall go the
opposite route, namely to $d=3$ spacetimes and investigate in the presence
of the Hoffmann term the impacts of the BI electrodynamics \cite{7}. This
suggests that beside its historical importance the BI and Hoffmann terms
together constitute viable a physical model. Our Lagrangian is shown (in the
Appendix) to satisfy the weak (WEC) and strong (SEC) energy conditions.
Gravity coupled BI electrodynamics in $2+1$-dimensions alone is known to be
singular \cite{8}. Some other forms of the non-linear electrodynamics
coupled with gravity in $2+1-$dimensions, also reveal that the possible
black hole solutions are singular \cite{8}.

Our model consists of the Einstein-Hilbert, cosmological $\left( \Lambda
\right) $ and BI terms which are supplemented by the logarithmic Hoffmann
term. As we had shown in our particle model \cite{4} the Hoffmann term
serves to introduce a natural boundary term so that the spacetime can be
considered divided in two different regions. The inner part which is
singularity-free turns out to contain a uniform electric field in addition
to the cosmological constant overall which is nothing but the anti-de Sitter
spacetime. The outer part is an entirely different spacetime which matches
according to the Israel's junction conditions \cite{9} on the circle,
identified as the $2-$ dimensional Friedmann-Robertson-Walker (FRW) universe.

An important result that we obtain within the context of Einstein-Hoffmann-
Born-Infeld (EHBI) theory in $d=3$ is that we can construct regular, charged
black hole solutions. This particular point constitutes the main motivation
for this letter. Let us note that Cataldo and Garc\'{\i}a \cite{10} found a
regular, charged BTZ black hole without identifying an initial Lagrangian
for the nonlinear electromagnetism. Unlike the BI case in the absence of
such a Lagrangian we can't investigate its implication in a flat spacetime
limit as well as the underlying thermodynamic properties. For the latter
analysis we appeal to the $1+1$-dimensional dilaton gravity model within the
Kaluza-Klein formalism. In this sense our work can be considered as an
alternative, charged singularity-free construction of the BTZ black hole 
\cite{11}. As the charged BTZ black hole has singularity at $r=0$ , with the
BI and Hoffmann terms in the action, it eliminates the singularity to yield
a regular black hole solution. Expectedly, in proper limits the boundary
circle is removed and our spacetime reduces to the uncharged, non-singular
BTZ black hole. Although, it does not allow us to interpret such a solution
as a particle model it may serve as a toy model in $d=3$, which can be
extended to higher dimensions as a tool to eliminate naked singularities and
restore the cosmic conjecture hypothesis, for instance. Further, inclusion
of the Hoffmann term may constitute an attractive field theory model of
elementary particles.

Technically, by the choice of parameters of the theory (i.e. BI parameter $b$%
, electric charge $q$, mass $M$ and the cosmological constant $\Lambda =-%
\frac{3}{l^{2}}$) we can adjust the event horizon to be located inside /
outside the circle whose radius is at $r_{0}=\left\vert \frac{q}{b}%
\right\vert $ (see Eq. (7) below). For $r<$ $r_{0}$ the spacetime is regular
with negative scalar curvature $R=-\frac{2}{b^{2}}$. By numerical analysis
we obtain solutions with single / double (or no) horizons which are
displayed in Fig. 1a.

\section{ACTION, FIELD EQUATIONS AND\ SOLUTIONS}

Our $3-$dimensional action in Einstein-Hoffmann-Born-Infeld (EHBI) theory,
is given by ($c=\hslash =k_{B}=G=\frac{1}{4\pi \epsilon _{\circ }}=1$)%
\begin{equation}
S=\int dx^{3}\sqrt{-g}\left( \frac{R-2\Lambda }{16\pi }+\mathcal{L}\left(
F\right) \right) ,
\end{equation}%
in which $F=\frac{1}{4}F_{\mu \nu }F^{\mu \nu }$ and 
\begin{equation}
\mathcal{L}\left( F\right) =\frac{2b^{2}}{4\pi }\left( 1-\sqrt{1+\frac{2F}{%
b^{2}}}+\ln \left( \frac{1+\sqrt{1+\frac{2F}{b^{2}}}}{2}\right) \right) ,
\end{equation}%
in which $b$ is a real parameter (HBI-parameter). The additional $\ln \left( 
\frac{1+\sqrt{1+\frac{2F}{b^{2}}}}{2}\right) $ term is known as the Hoffmann
term which supplements the more familiar BI Lagrangian. Also the limit of
the Lagrangian once $b^{2}\rightarrow \infty $ yields the Einstein Maxwell
linear electrodynamics as%
\begin{equation}
\lim_{b\rightarrow \infty }\mathcal{L}\left( F\right) =-\frac{1}{4\pi }F
\end{equation}%
and in the other limit once $F=0$ it goes to GR limit. The metric ansaetze
are axially symmetric which are given by 
\begin{equation}
ds^{2}=\left\{ 
\begin{array}{cc}
-\tilde{f}\left( r\right) dt^{2}+\tilde{f}\left( r\right)
^{-1}dr^{2}+r_{0}^{2}d\varphi ^{2}, & r\leq r_{0} \\ 
-f\left( r\right) dt^{2}+f\left( r\right) ^{-1}dr^{2}+r^{2}d\varphi ^{2}, & 
r\geq r_{0}%
\end{array}%
\right. ,
\end{equation}%
in which $r_{0}$ is the constant boundary to be determined. We comment here
that these two spacetimes are glued at $r=r_{0}.$ The electric field
two-form is chosen as%
\begin{equation}
\mathbf{F}=E\left( r\right) dt\wedge dr.
\end{equation}%
and the nonlinear Maxwell equation implies%
\begin{gather}
d\left( \mathcal{L}_{F}\ {}^{\star }\mathbf{F}\right) =0 \\
\left( \mathcal{L}_{F}=\frac{\partial \mathcal{L}}{\partial F}\right)  
\notag
\end{gather}

Consistent solution in both regions for the electric field is given by 
\begin{equation}
E\left( r\right) =\left\{ 
\begin{array}{cc}
\frac{q}{r_{0}}=b, & r\leq r_{0} \\ 
\frac{2qr}{r_{0}^{2}+r^{2}}, & r\geq r_{0}%
\end{array}%
\right. ,
\end{equation}%
in which $q$ is the electric charge and $r_{0}=\left\vert \frac{q}{b}%
\right\vert $ is the radius of a circle dividing the regions. Let us also
add that for $r\leq r_{0},$ the square-root term vanishes and $\mathcal{L}$
reduces to a constant which may be added to the cosmological constant $%
\Lambda .$ That is, for $r\leq r_{0}$ we have the anti-de Sitter spacetime.
From the expression (6) it can be seen that the electric potential is

\begin{equation}
V\left( r\right) =\left\{ 
\begin{array}{cc}
-q\ln 2+b\left( r_{0}-r\right) , & r\leq r_{0} \\ 
-q\ln \left( 1+\frac{r^{2}}{r_{0}^{2}}\right) , & r\geq r_{0}%
\end{array}%
\right. .
\end{equation}%
It is seen that $E_{r}=0$ at $r=0.$ $D_{r}$ is also zero (and similar result
holds when $r\rightarrow \infty $) which means that $D_{r}$ in terms of $%
E_{r}$ is single valued.

The energy density $u\left( r\right) =\frac{1}{2}\epsilon \overrightarrow{E}.%
\overrightarrow{D}$ $\left( \epsilon =\frac{\partial \mathcal{L}}{\partial F}%
,\text{ }\overrightarrow{D}=\epsilon \overrightarrow{E}\text{ }\right) $ can
be integrated to find the total energy, which diverges for \ $r\rightarrow
\infty .$

Variation of the action with respect to $g_{\mu \nu }$ yields the Einstein
equations%
\begin{equation}
G_{\mu }^{\nu }+\Lambda \delta _{\mu }^{\nu }=8\pi T_{\mu }^{\nu },
\end{equation}%
where 
\begin{equation}
T_{\mu }^{\nu }=\delta _{\mu }^{\nu }\mathcal{L-}\left( F_{\mu \lambda }F^{\
\nu \lambda }\right) \mathcal{L}_{F},
\end{equation}%
and in a closed form it reads%
\begin{equation}
T_{\mu }^{\nu }=\text{diag}\left[ \mathcal{L}-2F\mathcal{L}_{F},\mathcal{L}%
-2F\mathcal{L}_{F},\mathcal{L}\right] .
\end{equation}%
For $r<r_{0}$ the only nonzero component of the Einstein tensor $G_{\mu
}^{\nu }$ is $G_{\varphi }^{\varphi }=\frac{1}{2}f^{\prime \prime }\left(
r\right) $ in which the $tt$ and $rr$ parts of the Einstein equations simply
read 
\begin{equation}
\Lambda =8\pi T_{0}^{0}=8\pi T_{r}^{r}=\mathcal{L}-2F\mathcal{L}%
_{F}=-4b^{2}\ln 2,
\end{equation}%
which clearly imposes $\Lambda =-4b^{2}\ln 2,$ or equivalently 
\begin{equation}
\frac{1}{\ell ^{2}}=4b^{2}\ln 2,
\end{equation}%
since $\frac{1}{\ell ^{2}}=-\Lambda ,$ and this will be used wherever
necessary. Let us note that the relation between the two constants $\Lambda $
and $b$ is not a choice but is a direct consequence of the Einstein
equations. The other $\varphi \varphi $ component of the Einstein equation
for $r<\,r_{0}$ admits a solution which takes the form%
\begin{equation}
\tilde{f}\left( r\right) =\left( \frac{4q^{2}}{r_{0}^{2}}\right)
r^{2}+C_{1}r+C_{2},\text{ \ }r\leq r_{0}.
\end{equation}%
On the other hand the Einstein equations for $r>\,r_{0}$ \ give the metric
function%
\begin{equation}
\begin{array}{lc}
f\left( r\right) =-M+\frac{r^{2}}{\ell ^{2}}-4q^{2}\left[ \frac{r^{2}}{%
r_{0}^{2}}\ln \left( 1+\frac{r_{0}^{2}}{r^{2}}\right) +\ln \left( \frac{%
r^{2}+r_{0}^{2}}{\alpha ^{2}+r_{0}^{2}}\right) -\frac{\alpha ^{2}}{r_{0}^{2}}%
\ln \left( 1+\frac{r_{0}^{2}}{\alpha ^{2}}\right) \right] , & r\geq r_{0}%
\end{array}%
\end{equation}%
in which $M$ is the ADM mass of the black hole and $\alpha >0$ is a constant
introduced for dimensional reasons and, $C_{1}$ and $C_{2}$ are integration
constants to be found below. Once more we note that

\begin{equation}
\lim_{b\rightarrow \infty }f\left( r\right) =f_{CBTZ}\left( r\right) =-M+%
\frac{r^{2}}{\ell ^{2}}-8q^{2}\ln \frac{r}{\alpha }.
\end{equation}%
which removes the inner region automatically since $r_{0}\rightarrow 0$ and
leaves us with the charged BTZ (CBTZ) black hole. The other limit is given
by $q=0,$ which is the non-rotating BTZ (NBTZ) black hole, namely 
\begin{equation}
\lim_{q\rightarrow 0}f\left( r\right) =f_{NBTZ}\left( r\right) =-M+\frac{%
r^{2}}{\ell ^{2}}.
\end{equation}

Now, the boundary surface $F(r)=r-r_{0}=0$ can be considered as a $2-$%
dimensional FRW circular spacetime i.e.,%
\begin{equation}
ds_{circle}^{2}=-d\tau ^{2}+a\left( \tau \right) ^{2}d\phi ^{2}
\end{equation}%
in which $\tau $ is the proper time on the circle. Furthermore, the two
metric functions in (14) and (15) must fulfill the Israel junction
conditions \cite{9,12,13} reading%
\begin{equation}
S_{i}^{j}=-\frac{1}{8\pi }\left( \left\langle K_{i}^{j}\right\rangle
-\left\langle K\right\rangle \delta _{i}^{j}\right) ,
\end{equation}%
where $S_{i}^{j}$ is the stress energy tensor on the boundary, $K_{i}^{j}$
is the extrinsic curvature, $K$ is the trace of it and $\left\langle
K_{i}^{j}\right\rangle =\left( K_{i}^{j}\right) _{r_{\circ }^{+}}-\left(
K_{i}^{j}\right) _{r_{\circ }^{-}}.$ One can show that the Israel equations
read%
\begin{eqnarray}
S_{\tau }^{\tau } &=&\frac{1}{8\pi }\left( \frac{\sqrt{f\left( a\right) }}{a}%
-\frac{\sqrt{\tilde{f}\left( a\right) }}{a}\right) _{a=r_{0}} \\
S_{\varphi }^{\varphi } &=&\frac{1}{16\pi }\left( \frac{f^{\prime }\left(
a\right) }{\sqrt{f\left( a\right) }}-\frac{\tilde{f}^{\prime }\left(
a\right) }{\sqrt{\tilde{f}\left( a\right) }}\right) _{a=r_{0}},
\end{eqnarray}%
where a prime is a derivative with respect to $a$. A smooth transition from $%
r\geq r_{0}$ into $r\leq r_{0}$ requires that $S_{\tau }^{\tau }=S_{\phi
}^{\phi }=0$ which are equivalent to both, continuous metric and its first
derivatives at $r=r_{0}.$ These conditions fix the constants as 
\begin{eqnarray}
C_{1} &=&-\frac{8q^{2}}{r_{0}}, \\
C_{2} &=&-M+4q^{2}\left[ 1-\frac{\alpha ^{2}}{r_{0}^{2}}\ln \left( \frac{%
\alpha ^{2}}{\alpha ^{2}+r_{0}^{2}}\right) -\ln \left( \frac{2r_{0}^{2}}{%
\alpha ^{2}+r_{0}^{2}}\right) \right] .
\end{eqnarray}%
The Ricci and Kretschmann scalars for $r\leq r_{0}$ are

\begin{equation}
R=-8b^{2},\text{ \ \ \ }K=64b^{4}
\end{equation}%
so that the inner space time is regular as long as $b<\infty .$

From (14) and (15) it can be shown easily that at the boundary $r=r_{0}$ we
have the conditions $\tilde{f}(r_{0})=f(r_{0})$ and $\tilde{f}^{\prime
}(r_{0})=f^{\prime }(r_{0})=0$, satisfied. For chosen parameters $q$, $%
\alpha ,$ $M$ and $b$ our metric gives rise, through numerical analysis to
different black holes (see Fig. 1a). The interesting case is the one with
event horizon from $\tilde{f}(r)=0$ and an inner horizon from $f(r)=0$ (Fig.
1a C). The fact that our darkness function for the common parameters
satisfies black hole condition, both outside and inside, can be seen from
the numerical plots.

\subsection{Tidal force in the inner region $r\leq r_{0}$}

As we found at the end of the previous section, the origin is nonsingular.
It should be noted that the tidal force may reveal a singularity even when
the curvature invariants are finite.

For checking this we use the formalism introduced in Ref. \cite{14}, by
considering the line element in the form%
\begin{equation}
ds^{2}=-\frac{F\left( r\right) }{G\left( r\right) }dt^{2}+\frac{dr^{2}}{%
F\left( r\right) }+R\left( r\right) ^{2}d\varphi ^{2},
\end{equation}%
in which $F\left( r\right) =f\left( r\right) ,$ $G\left( r\right) =1$ and $%
R\left( r\right) =r_{0}$ for our case. The only nonzero component of the
curvature tensor in the static observer's orthonormal basis is given by%
\begin{equation}
R_{trtr}=\frac{1}{2}f^{\prime \prime }\left( r\right)
\end{equation}%
and those related by the symmetry. An observer who is radially falling
freely toward the center with conserved energy $E$ is related to the static
orthonormal frame by a local Lorantz boost \cite{15}. The instantaneous
velocity of the observer is given by%
\begin{equation}
v=\sqrt{1-\frac{F}{GE^{2}}}=\sqrt{1-\frac{f\left( r\right) }{E^{2}}}
\end{equation}%
and the only nonzero curvature component in the Lorantz boosted frame
happens to be%
\begin{equation}
R_{\hat{t}\hat{r}\hat{t}\hat{r}}=R_{trtr}=\frac{1}{2}f^{\prime \prime
}\left( r\right) =4b^{2}.
\end{equation}%
Having the other Riemann components zero means that the tidal force in the
transverse direction vanishes. This could be also seen directly from the Eq.
(2.9) in Ref. \cite{14}.

\section{\textbf{Thermodynamical considerations through the }1+1-dimensional
dilaton gravity}

In this section we follow \cite{16} to study the thermodynamics of the EHI
black hole found above. To do so we derive the solution (15) for $r>r_{0}$
from the dilaton gravity. Now we consider 
\begin{equation}
ds^{2}=g_{ab}dx^{a}dx^{b}=\tilde{g}_{\mu \nu }d\tilde{x}^{\mu }d\tilde{x}%
^{\nu }+\phi ^{2}\left( \tilde{x}\right) d\theta ^{2}
\end{equation}%
in which $\phi $ is the radius of the circle $S^{1}$ in $M_{3}=M_{2}\times
S^{1}.$ Herein the Greek indices represent the two-dimensional spacetime.
After the Kaluza-Klein dimensional reduction, the action (1) becomes%
\begin{equation}
S_{2D}=2\pi \int d\tilde{x}^{2}\sqrt{-\tilde{g}}\phi \left( \frac{\tilde{R}%
-2\Lambda }{16\pi }+\mathcal{L}\left( F\right) \right) 
\end{equation}%
in which $\tilde{R}$ is the Ricci scalar of $M_{2}$ and for the sake of
completeness we shall describe the dilatonic approach briefly. The field
equations are given by%
\begin{eqnarray}
d\left( \phi \mathcal{L}_{F}\ {}^{\star }\mathbf{F}\right)  &=&0, \\
\nabla ^{2}\phi +2\phi \Lambda  &=&16\pi \phi \left( \mathcal{L}\left(
F\right) -2F\mathcal{L}_{F}\right) , \\
\tilde{R}-2\Lambda  &=&-16\pi \mathcal{L}\left( F\right) 
\end{eqnarray}%
in which $\mathbf{F=}E\left( \phi \right) dt\wedge d\phi $ and ${}^{\star }%
\mathbf{F=}E\left( \phi \right) $ (i.e., $0-$form). Our choice of electric
field yields%
\begin{equation}
F=-\frac{1}{2}E\left( \phi \right) ^{2}
\end{equation}%
and therefore the Maxwell equation (31) implies%
\begin{equation}
\frac{\phi E\left( \phi \right) }{1+\sqrt{1-\frac{E\left( \phi \right) ^{2}}{%
b^{2}}}}\ {}\mathbf{=}q=\text{constant}
\end{equation}%
which admits%
\begin{equation}
E\left( \phi \right) =\frac{2q\phi }{q^{2}b^{2}+\phi ^{2}}.
\end{equation}%
Accordingly the Lagrangian $\mathcal{L}\left( F\right) $ takes the form 
\begin{equation}
\mathcal{L}\left( F\right) =\frac{b^{2}}{2\pi }\left[ \frac{2q^{2}}{%
q^{2}+b^{2}\phi ^{2}}+\ln \left( \frac{b^{2}\phi ^{2}}{q^{2}+b^{2}\phi ^{2}}%
\right) \right] 
\end{equation}%
with%
\begin{equation}
\mathcal{L}_{F}=-\frac{q^{2}+b^{2}\phi ^{2}}{4\pi b^{2}\phi ^{2}}.
\end{equation}%
The other two equations also become%
\begin{equation}
\nabla ^{2}\phi =V\left( \phi \right) =\frac{2\phi }{\ell ^{2}}+8\phi
b^{2}\ln \left( \frac{b^{2}\phi ^{2}}{q^{2}+b^{2}\phi ^{2}}\right) 
\end{equation}%
and 
\begin{equation}
\tilde{R}=-V^{\prime }\left( \phi \right) =-\frac{2\phi }{\ell ^{2}}-\frac{%
16q^{2}b^{2}}{q^{2}+b^{2}\phi ^{2}}-8b^{2}\ln \left( \frac{b^{2}\phi ^{2}}{%
q^{2}+b^{2}\phi ^{2}}\right) .
\end{equation}%
Here for our future use we find the asymptotical behavior of the potential $%
V\left( \phi \right) $ and its first derivative $V^{\prime }\left( \phi
\right) $ which are given by%
\begin{eqnarray}
V_{_{CBTZ}}\left( \phi \right)  &=&\lim_{b\rightarrow \infty }V\left( \phi
\right) =\frac{2\phi }{\ell ^{2}}-\frac{8q^{2}}{\phi },\text{ } \\
V_{_{CBTZ}}^{\prime }\left( \phi \right)  &=&\lim_{b\rightarrow \infty
}V^{\prime }\left( \phi \right) =\frac{2\phi }{\ell ^{2}}+\frac{8q^{2}}{\phi 
}.
\end{eqnarray}%
These two latter equations correspond to the $2-$dimensional field equations
of dilaton gravity with an action%
\begin{equation}
S_{2D}=\int_{M_{2}}d\tilde{x}dt\sqrt{-\tilde{g}}\left( \phi \tilde{R}%
+V\left( \phi \right) \right) 
\end{equation}%
and the line element 
\begin{equation}
ds^{2}=-f\left( \tilde{x}\right) dt^{2}+\frac{d\tilde{x}^{2}}{f\left( \tilde{%
x}\right) }.
\end{equation}%
The field equations simply read%
\begin{equation}
\nabla ^{2}\phi \left( \tilde{x}\right) =f\left( \tilde{x}\right) \phi
^{\prime \prime }\left( \tilde{x}\right) +f^{\prime }\left( \tilde{x}\right)
\phi ^{\prime }\left( \tilde{x}\right) =V\left( \phi \right) 
\end{equation}%
(in which a prime means derivative with respect to the argument) and 
\begin{equation}
\tilde{R}=-f^{\prime \prime }\left( \tilde{x}\right) =-V^{\prime }\left(
\phi \right) .
\end{equation}%
For a linear dilaton ansatz which we impose at this stage 
\begin{equation}
\phi \left( \tilde{x}\right) =\tilde{x}
\end{equation}%
it simply yields from (45) that%
\begin{equation}
f^{\prime }\left( \tilde{x}\right) =V\left( \phi \right) .
\end{equation}%
Now, we consider the results found above with 
\begin{equation}
ds^{2}=-f\left( \phi \right) dt^{2}+\frac{d\phi ^{2}}{f\left( \phi \right) }
\end{equation}%
where $f\left( \phi \right) $ is aptly expressed by 
\begin{equation}
f\left( \phi \right) =J\left( \phi \right) -\mathcal{C}
\end{equation}%
in which $J\left( \phi \right) $ is defined as 
\begin{equation}
J\left( \phi \right) =\int V\left( \phi \right) d\phi =\frac{\phi ^{2}}{\ell
^{2}}-4q^{2}\ln \left( \frac{q^{2}+b^{2}\phi ^{2}}{q^{2}+b^{2}\phi _{0}^{2}}%
\right) +4b^{2}\phi ^{2}\ln \left( \frac{b^{2}\phi ^{2}}{q^{2}+b^{2}\phi ^{2}%
}\right) -4b^{2}\phi _{0}^{2}\ln \left( \frac{b^{2}\phi _{0}^{2}}{%
q^{2}+b^{2}\phi _{0}^{2}}\right) .
\end{equation}%
Let us comment at this point that any choice other than the linear dilaton
(47), may not coincide with the $2+1-$dimensional solution. With reference
to \cite{16} $\mathcal{C}$ represents the ADM mass of the EHBI black hole
(See the Appendix of Ref. \cite{16}). Herein $\phi _{0}$ is just a reference
potential which is set to be $\phi _{0}=\alpha =\ell $ with $\Lambda =-\frac{%
1}{\ell ^{2}}=-4b^{2}\ln 2.$ Following \cite{16}, the extremal value for $%
\phi _{+}$ is found by $V\left( \phi _{+}=\phi _{e}\right) =0,$ which is
given by%
\begin{equation}
\phi _{e}^{2}=\frac{q^{2}}{b^{2}}=r_{0}^{2}
\end{equation}%
leading us to an extremal mass as%
\begin{equation}
M_{e}=\left( \frac{1+4q^{2}\ln 2}{\ln 2}\right) \ln \left( 1+4q^{2}\ln
2\right) -4q^{2}\ln \left( 8q^{2}\ln 2\right) .
\end{equation}%
By trading the Maxwell terms for a dilaton potential, we now have the charge
parameter $q$ appearing in the action. Efectively, this restricts our
analysis to the fixed charge (canonical) ensemble, as that parameter can no
longer be varied \cite{17}. To have at least one horizon one should set $%
M\geq M_{e}$ so that $\phi _{+}$ indicates the outer horizon at which the
Hawking temperature is defined by%
\begin{equation}
T_{H}=\frac{V\left( \phi _{+}\right) }{4\pi }=\frac{2b^{2}\phi _{+}}{\pi }%
\ln \left( \frac{2b^{2}\phi _{+}^{2}}{q^{2}+b^{2}\phi _{+}^{2}}\right) .
\end{equation}

The heat capacity and free energy are given respectively by%
\begin{equation}
C_{Q}\left( \phi _{+}\right) =4\pi \left( \frac{V\left( \phi _{+}\right) }{%
V^{\prime }\left( \phi _{+}\right) }\right) =\frac{4\pi \phi _{+}\ln \left( 
\frac{2b^{2}\phi _{+}^{2}}{q^{2}+b^{2}\phi _{+}^{2}}\right) }{\frac{2q^{2}}{%
q^{2}+b^{2}\phi _{+}^{2}}+\ln \left( \frac{2b^{2}\phi _{+}^{2}}{%
q^{2}+b^{2}\phi _{+}^{2}}\right) }
\end{equation}%
and%
\begin{equation}
F\left( \phi _{+}\right) =J\left( \phi _{+}\right) -J\left( \phi _{e}\right)
-\phi _{+}V\left( \phi _{+}\right) =4q^{2}\ln \left( \frac{2q^{2}}{%
q^{2}+b^{2}\phi _{+}^{2}}\right) -4b^{2}\phi _{+}^{2}\ln \left( \frac{%
2b^{2}\phi _{+}^{2}}{q^{2}+b^{2}\phi _{+}^{2}}\right) ,
\end{equation}%
which complete the thermodynamical analysis of our EHBI black hole. We note
that, in this stage we can not take the limit $b\rightarrow \infty $ (which
is CBTZ limit) because we have set some of our constants in terms of each
other to satisfy the Israel boundary conditions. Instead, to use these
results we may keep our parameters as they appeared before our conditions
are imposed and then our results provide the correct CBTZ limit i.e., 
\begin{equation}
T_{H}^{\left( CBTZ\right) }=\lim_{b\rightarrow \infty }\frac{V\left( \phi
_{+}\right) }{4\pi }=\frac{1}{4\pi }\left( \frac{2\phi _{+}}{\ell ^{2}}-%
\frac{8q^{2}}{\phi _{+}}\right) ,
\end{equation}%
and the same for the other cases. Nevertheless with the limit $b\rightarrow
\infty $ (i.e. $r_{0}\rightarrow 0$) it automatically removes the inner
space. This leads to the case that we don't need to consider any boundary
conditions and therefore the relation between $\ell $ and $b$ will no longer
be needed. Under these circumstances we can use the above limits.

Through the explicit expressions of $T_{H}$, $C_{Q}\left( \phi _{+}\right) $
and $F\left( \phi _{+}\right) $ one simply finds that by a rescaling $%
b^{2}\phi _{+}^{2}\rightarrow \tilde{\phi}_{+}^{2}$ one can eliminate $b$
from the expressions. This means that the different values for $b$ (with
fixed $q)$ does not change the general behavior of the thermodynamical
properties. In such case we set $b=1$ and plot Fig.s 1 and 2 without loss of
generality of the problem. Fig. 1a displays through the Israel's junction
conditions that there are three possible gluing: i) $M<M_{e}$ corresponds to
no horizon, ii) $M=M_{e}$ corresponds to a single horizon and finally iii) $%
M>M_{e}$ corresponds to two distinct horizons in which $r_{-}<r_{0}<r_{+}.$
Note also that $r=r_{0}$ is the only minimum point of the metric (see Fig.
1a). Fig. 1b is a plot of extremal mass $M_{e}$ versus charge $q.$ This
reveals the regions in which the mass $M$ must be chosen in terms of $q$ to
have two horizons.

Fig. 2 displays the thermodynamical quantities versus horizon. Let us add
that the only feasible case we are interested is for $\phi _{+}>r_{0}.$
Fig.s 2a, 2b and 2c confirm that the spacetime is thermodynamically stable.

\section{Conclusion}

The objective of this study was to construct an electrically charged regular
black hole solution as an extension of the uncharged BTZ black hole. Within
the familiar linear Maxwell electromagnetism this is not available. Next
attempt naturally is to consider nonlinear electrodynamics as a potential
candidate that may host such a black hole. Now, with the additional
Hoffmannn term in the BI Lagrangian an alternative method of gluing
different spacetimes renders construction of regular black holes possible.
As explained, the role of the Hoffmann term, is to introduce a natural
circle as boundary with radius $r_{0}=\left\vert \frac{q}{b}\right\vert $\
(with $q=$electric charge and $b=$BI-parameter) which divides spacetime into
inside ($r<r_{0}$) and outside ($r>r_{0}$)\ regions. In the process of
matching of spacetimes at the junction we employ Israel's junction
conditions. The inner region is nothing but the $3$-dimensional anti-de
Sitter spacetime in which the cosmological constant is proportional to the
BI parameter. It should also be added that there is no lower bound on $r_{0},
$ that is, $r_{0}\rightarrow 0$ with $b\rightarrow \infty $, which removes
the inner region, leaving behind the singular, charged BTZ spacetime. We
obtained a variety of black hole states as depicted in Fig. 1a. From the
thermodynamical requirements (i.e. $T_{H}>0$) we single out the class that
are constrained by the feasibility conditions. This means automatically that
occurrence of horizons outside and on the circle with radius $r_{0}$ become
the only physical cases. Finally the thermodynamics of constructed EHBI
black hole has been analyzed in accordance with the Kaluza-Klein reduction
to $1+1$-dimensional dilaton gravity. The relevant thermodynamical
quantities are depicted in Fig. 2.

\textbf{Figure captions}

Fig. 1a: Plot of the metric function $f(r)$ versus $r$ for different masses $%
M$, expressed in terms of the extremal mass $M_{e}.$ Although we analyzed
the case $r\geq r_{0}$ the continuity requirements through Israel junction
conditions demand also to cover the region $r\leq r_{0}$. Three different
cases have been shown: double horizons ($M=2M_{e}$) single horizon ($M=M_{e}$%
) and no horizons ($M=0.5M_{e}$). We note that from (13) in the text we have
chosen $\ell ^{2}=\frac{1}{4b^{2}\ln 2}$ and also $\alpha =\ell .$

Fig. 1b. Extremal mass $M_{e}$ versus the charge $q$.

Fig. 2: The thermodynamical quantities for the EHBI black hole. (a) The
Hawking temperature $T_{H}\left( \phi _{+}\right) $, which is positive for $%
\phi _{+}\geq 1$ (the event horizon). (b) The specific heat $C\left( \phi
_{+}\right) $, behaves also well for $\phi _{+}\geq 1=r_{0}.$ The phase
change occurs for $\phi _{+}\leq 1=r_{0}$ which will clearly lie in the
infeasible region. (c) The free energy $F\left( \phi _{+}\right) $ which is
negative for the $\phi _{+}\geq 1.$

\textbf{APPENDIX: Energy Conditions}

Our energy momentum tensors in two regions are given by:%
\begin{equation}
T_{\mu }^{\nu }=\left\{ 
\begin{array}{ll}
\text{diag}\left[ -\frac{q^{2}}{2\pi r_{0}^{2}}\ln \left( 1+\frac{r_{0}^{2}}{%
r^{2}}\right) ,-\frac{q^{2}}{2\pi r_{0}^{2}}\ln \left( 1+\frac{r_{0}^{2}}{%
r^{2}}\right) ,-\frac{q^{2}}{2\pi r_{0}^{2}}\ln \left( 1+\frac{r_{0}^{2}}{%
r^{2}}\right) +\frac{q^{2}}{\pi \left( r_{0}^{2}+r^{2}\right) }\right] , & 
r\geq r_{0} \\ 
\text{diag}\left[ -\frac{q^{2}}{2\pi r_{0}^{2}}\ln \left( 2\right) ,-\frac{%
q^{2}}{2\pi r_{0}^{2}}\ln \left( 2\right) ,-\frac{q^{2}}{2\pi r_{0}^{2}}\ln
\left( 2\right) +\frac{q^{2}}{2\pi r_{0}^{2}}\right] , & r\leq r_{0}%
\end{array}%
\right. .  \tag{A1}
\end{equation}%
Accordingly one finds the energy density $\rho $ and the principal pressures 
$p_{i}$ as%
\begin{eqnarray}
\rho  &=&-T_{t}^{t}=\left\{ 
\begin{array}{ll}
\frac{q^{2}}{2\pi r_{0}^{2}}\ln \left( 1+\frac{r_{0}^{2}}{r^{2}}\right) , & 
r\geq r_{0} \\ 
\frac{q^{2}}{2\pi r_{0}^{2}}\ln \left( 2\right) , & r\leq r_{0}%
\end{array}%
\right. ,\text{ \ \ \ \ \ \ \ \ }p_{r}=T_{r}^{r}=\left\{ 
\begin{array}{ll}
-\frac{q^{2}}{2\pi r_{0}^{2}}\ln \left( 1+\frac{r_{0}^{2}}{r^{2}}\right) , & 
r\geq r_{0} \\ 
-\frac{q^{2}}{2\pi r_{0}^{2}}\ln \left( 2\right) , & r\leq r_{0}%
\end{array}%
\right. ,\text{ \ \ \ \ \ }  \TCItag{A2} \\
\text{\ \ \ }p_{\varphi } &=&T_{\varphi }^{\varphi }==\left\{ 
\begin{array}{ll}
-\frac{q^{2}}{2\pi r_{0}^{2}}\ln \left( 1+\frac{r_{0}^{2}}{r^{2}}\right) +%
\frac{q^{2}}{\pi \left( r_{0}^{2}+r^{2}\right) }, & r\geq r_{0} \\ 
-\frac{q^{2}}{2\pi r_{0}^{2}}\ln \left( 2\right) +\frac{q^{2}}{2\pi r_{0}^{2}%
}, & r\leq r_{0}%
\end{array}%
\right. .  \notag
\end{eqnarray}

\subsection{Weak Energy Condition (WEC)}

The WEC states that,

\begin{equation}
\rho \geq 0\text{ \ \ \ \ \ \ \ \ \ \ \ and \ \ \ \ \ \ \ \ }\rho +p_{i}\geq
0\text{ \ \ \ \ \ }(i=1,2)  \tag{A3}
\end{equation}%
which are satisfied in both regions.

\subsection{Strong Energy Condition (SEC)}

This condition states that;

\begin{equation}
\rho +\dsum\limits_{i=1}^{2}p_{i}\geq 0\text{ \ \ \ \ \ \ and \ \ \ \ \ \ }%
\rho +p_{i}\geq 0,  \tag{A4}
\end{equation}%
which are also satisfied.

\bigskip 

\bigskip 

\end{document}